\documentclass[aps,twocolumn,showpacs,superscriptaddress,preprintnumbers,floatfix,amsmath,amssymb,eufrak,table]{revtex4}
\usepackage{ulem}
\usepackage{color}
\usepackage{graphicx}  
\usepackage{appendix}
 \usepackage{epsfig}
\usepackage[colorlinks=true,linkcolor= red, citecolor = cyan, urlcolor = blue ]{hyperref}
\usepackage{epstopdf}
\usepackage{amsmath}
\usepackage{subfig}
\usepackage{comment}
\usepackage{float}
\usepackage{multirow}
\usepackage{ragged2e} 

\newcommand{\be}{\begin{equation}}
	\newcommand{\ee}{\end{equation}}
\newcommand{\ba}{\begin{eqnarray}}
	\newcommand{\ea}{\end{eqnarray}}

\begin{document}
\title{Quarkonium $p_{\rm T}$ spectra in heavy--ion collisions at LHC energies within a hydrodynamic core--corona framework}
	
\author{Biswarup Paul}\email{biswarup.babu@gmail.com}
\affiliation{Bali Ram Bhagat College, Samastipur - 848101, Bihar, India}

\date{\today}
\begin{abstract}
We present a systematic study of the transverse momentum ($p_{\rm T}$) spectra of charmonium (J/$\psi$, $\psi(2S)$) and bottomonium ($\Upsilon(nS)$) states in Pb–Pb collisions at $\sqrt{s_{\rm NN}} = 5.02$ TeV within an analytical relativistic hydrodynamics framework. The medium evolution is described assuming cylindrical symmetry with boost-invariant longitudinal expansion and Hubble-like transverse flow. Quarkonium spectra are evaluated using the Cooper–Frye formalism on a constant-temperature freeze-out hypersurface, supplemented by a core–corona approach to include both thermal and non-thermal contributions. The model describes the measurements from ALICE and CMS over a broad $p_{\rm T}$ range. For charmonium, both the spectra and the $\psi(2S)$/J/$\psi$ ratio are well reproduced, while deviations at high $p_{\rm T}$ for J/$\psi$ indicate additional hard production mechanisms. In the bottomonium sector, the $\Upsilon(nS)$ spectra and their yield ratios are successfully described, consistent with the expected sequential suppression pattern. These results demonstrate that an analytical hydrodynamic approach combined with a core–corona framework provides a unified and transparent description of quarkonium production in heavy--ion collisions at LHC energies.
\end{abstract}
\pacs{Quarkonium, QGP, Relativistic Hydrodynamics, LHC}
\keywords{Relativistic heavy-ion collisions, Quark Gluon Plasma, Quarkonium, Relativistic Hydrodynamics, LHC.}
	
\maketitle

\section{Introduction}

Quarkonia, the bound states of a heavy quark ($Q$) and its antiquark ($\bar{Q}$), constitute one of the most sensitive probes of the hot and dense medium created in relativistic nucleus--nucleus ($A$--$A$) collisions. Charmonium ($c\bar{c}$) and bottomonium ($b\bar{b}$) states are particularly well suited for this purpose due to their large masses and formation at early stages of the collision. In the presence of a deconfined quark--gluon plasma (QGP), the color interaction between heavy quarks is screened by the surrounding medium, analogous to Debye screening in electromagnetic plasmas~\cite{Matsui}. This leads to the dissociation of quarkonium states and consequently to a suppression of their yields in heavy--ion collisions relative to proton--proton ($pp$) and proton--nucleus ($p$--$A$) collisions~\cite{Rvogt,Satz,Kluberg,Rapp,Zhao}. In addition to color screening, in-medium interactions with thermal partons introduce an imaginary component to the heavy-quark potential, resulting in a finite thermal width and eventual dissociation of quarkonium states~\cite{Mocsy,Rothkopf}. Since these effects depend on both the medium temperature and the binding energy of the state, quarkonium suppression follows a characteristic sequential pattern, where loosely bound states melt at lower temperatures than more tightly bound ones~\cite{Karsch,Bhaduri}. This sequential suppression provides a direct handle on the temperature evolution and screening properties of the QGP.

Experimentally, quarkonium production has been extensively investigated at the CERN SPS~\cite{Abreu,Arnaldi}, RHIC~\cite{Adare1,Adare2,Adam}, and at the CERN LHC~\cite{Abelev,Adam2,Adam3,Alice_quarkonia,ALICE_PsiP_PRL,ALICE_JPsi_midy,CMS,CMS_PsiP,CMS_Prompt,LHCb_quarkonia,Atlas_quarkonia,ALICE_Y1S_fwdy,CMS_Y1S_midy,CMS_Y2S_midy,CMS_Y3S_midy,ATLAS_YnS_midy} in heavy-ion collisions.
The modification of yields is commonly quantified through the nuclear modification factor $R_{AA}$, which encodes the combined effects of suppression and regeneration. At LHC energies, the observed weaker suppression of J/$\psi$ at low $p_{\rm T}$, compared to lower energies at RHIC, indicates a substantial contribution from regeneration via recombination of deconfined charm quarks. The $p_{\rm T}$ and rapidity dependence of $R_{AA}$ further highlight the interplay between suppression and regeneration in a medium with high heavy-quark densities. In contrast, bottomonium production is largely governed by suppression effects, as the significantly smaller $b\bar{b}$ production cross section renders regeneration negligible. Measurements by the CMS~\cite{CMS_Y1S_midy,CMS_Y2S_midy,CMS_Y3S_midy} and ATLAS~\cite{ATLAS_YnS_midy} reveal a clear hierarchy in the suppression of $\Upsilon(1S)$, $\Upsilon(2S)$, and $\Upsilon(3S)$ states, consistent with their binding energies and the expected sequential melting pattern~\cite{Du}. This makes bottomonia particularly clean probes of the early, hottest stages of the medium, where color screening dominates.

While the nuclear modification factor $R_{AA}$ is useful for identifying medium effects, it is not optimal for probing thermal properties. Instead, transverse momentum spectra and elliptic flow ($v_2$) are more sensitive to thermalization and collective behavior. The measured $p_{\rm T}$ spectra are often analyzed within hydrodynamics-inspired blast-wave models to assess the extent to which heavy quarks thermalize and participate in the medium expansion. Early evidence of thermal features in charmonium production was reported in Ref.~\cite{Gazdzicki} through the analysis of J/$\psi$ ($\psi'$) $p_{\rm T}$ spectra in $\sqrt{s_{\rm NN}} = 17.3$ GeV Pb--Pb collisions at SPS. Similar analyses within boosted thermal model framework have been performed for RHIC~\cite{Bugaev} and LHC~\cite{Anton_SHM} data, assuming coincident kinetic and chemical freeze-out due to negligible hadronic rescattering. For bottomonia, $\Upsilon$ $p_{\rm T}$ spectra measured by CMS at $\sqrt{s_{\rm NN}} = 2.76$ TeV~\cite{CMS_YnS} remain inconclusive regarding blast-wave descriptions~\cite{Reygers}. At $\sqrt{s_{\rm NN}} = 5.02$ TeV, measurements by CMS and ALICE show that $\Upsilon(1S)$ exhibits significantly smaller $v_2$ than inclusive J/$\psi$~\cite{ALICE_v2_JPsi,CMS_v2_JPsi,ALICE_v2_JPsi2,ALICE_v2_JPsi3,ALICE_v2_JPsi4}. While the $v_2(p_T)$ of J/$\psi$ can be described by transport models including regeneration~\cite{Rapp_JPsi}, the $\Upsilon$ $v_2$ is consistent with zero within uncertainties. These observations can be explained both by models neglecting $b$-quark thermalization~\cite{Du,Bhaduri_Y,Bhaduri_Y2} and by blast-wave scenarios assuming collective flow~\cite{Reygers}, leaving the degree of bottom-quark thermalization inconclusive.


An alternative description of quarkonium production is provided by the statistical hadronization model (SHM), in which all heavy quark pairs are produced in initial hard scatterings and subsequently hadronize at the QCD phase boundary according to statistical weights~\cite{Anton_SHM2,Anton_SHM3,Anton_SHM4,Anton_SHM5}. In this framework, quarkonia are assumed to undergo chemical freeze-out at a temperature similar to that of light hadrons ($T_f \approx 156$ MeV). However, analyses based on relative quarkonium yields suggest a different scenario. Early studies indicated a significantly higher freeze-out temperature for bottomonium, $T_f = 222^{+28}_{-29}$ MeV~\cite{Gupta}, motivating a flavor-dependent freeze-out picture~\cite{Gupta2}. More recent investigations show that bottomonium states may freeze out at $T_f \sim 230$ MeV~\cite{Bhaduri_T}, largely independent of collision energy, rapidity, and centrality, consistent with approximate boost invariance and early thermalization. In the charmonium sector, the situation is more complex. While analyses of ALICE data at low $p_{\rm T}$ support early thermalization, the interpretation of CMS measurements is limited by high $p_{\rm T}$ thresholds. Moreover, extensions of thermal analyses to smaller systems such as $p$--Pb and high-multiplicity $pp$ collisions reveal qualitatively similar trends but yield unrealistically large freeze-out temperatures, indicating the importance of non-equilibrium effects. These observations suggest that heavy quarkonia may decouple from the medium at temperatures significantly higher than light hadrons, pointing to a nontrivial interplay between thermal and non-thermal production mechanisms.

From a theoretical perspective, understanding quarkonium production requires a consistent description of both the medium evolution and particle emission. While transport models incorporate dissociation and regeneration dynamically, hydrodynamic approaches provide a macroscopic description of the medium evolution. In this context, analytical solutions of relativistic hydrodynamics play a crucial role in establishing benchmarks for more sophisticated numerical simulations~\cite{Hydro2,Hydro3}. Although general solutions are difficult to obtain, simplified configurations with specific symmetries allow for exact analytical treatments. 
Recently, analytical hydrodynamic solutions for systems with spherical and cylindrical symmetry, boost-invariant longitudinal expansion, and Hubble-like transverse flow have been developed~\cite{Hydro_core}. These solutions yield closed-form expressions for the temperature evolution as a function of proper time and enable a consistent construction of the freeze-out hypersurface. Importantly, they lead to exact analytical expressions for particle momentum distributions via the Cooper--Frye prescription. 
Furthermore, analytical expressions for average radial and transverse flow velocities can be derived directly from the hydrodynamic profiles, providing a clear physical interpretation of the model parameters.

In this work, we employ the analytical hydrodynamic framework developed in Ref~\cite{Hydro_core} in which the medium evolution is described by cylindrical symmetry with boost-invariant longitudinal expansion and Hubble-like transverse flow, appropriate for LHC energies, to study the transverse momentum spectra of quarkonium states in Pb--Pb collisions at $\sqrt{s_{\rm NN}} = 5.02$ TeV. 
To incorporate both equilibrated and non-equilibrated contributions, we adopt a core--corona approach~\cite{GM,Anton_SHM}, where the core represents the thermalized medium undergoing collective expansion calculated from the analytical hydrodynamic framework, while the corona accounts for quarkonia produced in initial hard scatterings. The corona contribution is modeled using a power-law functional form motivated by transverse momentum spectra measured in $pp$ collisions at the same collision energy. The parameters of this contribution are constrained by experimental $p_{\rm T}$ distributions and reflect the characteristic behavior of hard processes. The boundary between the core and corona is determined from the nuclear charge density distribution, corresponding to a threshold where the local density drops to approximately 10\% of its central value. The relative fractions of the core and corona components are evaluated using a Glauber Monte Carlo approach~\cite{GM,Anton_SHM}.



The paper is organized as follows. Section II outlines the theoretical framework based on analytical hydrodynamic solutions, supplemented by a core–corona description to account for both thermal and non-thermal contributions. Section III presents the results and their comparison with experimental data, followed
by the summary and conclusions in Section IV.

\section{Theoretical formalism}

We employ an exact analytical solution of relativistic hydrodynamics developed in Ref.~\cite{Hydro_core} to describe the space–time evolution of the medium formed in Pb--Pb collisions at $\sqrt{s_{\rm NN}} = 5.02$ TeV. The framework assumes cylindrical symmetry with boost-invariant longitudinal expansion and Hubble-like transverse flow, appropriate for ultra-relativistic energies. The $p_{\rm T}$ spectra are evaluated on a constant-temperature freeze-out hypersurface using the Cooper--Frye prescription. The formulation is based on ideal hydrodynamics with a simplified equation of state characterized by a constant speed of sound, $c_s$. Although $c_s$ does not appear explicitly in the final expression for the particle spectra, it governs the temperature evolution and its proper-time dependence. As a result, the freeze-out hypersurface, defined by a fixed temperature and corresponding proper time, is implicitly determined by the equation of state. In this way, the influence of $c_s$ remains encoded in the final particle distributions. The results should therefore be interpreted within this approximation. Within this framework, the transverse momentum distribution of hadrons emitted from the expanding medium can be obtained in a closed analytical form~\cite{Hydro_core}, given by
\begin{align}
E \frac{dN}{d^3p}
&=
\frac{g R^3}{2\pi^2}
\sum_{n=1}^{\infty} \epsilon_n
\int_{0}^{1}
\Bigg[
m_T \sqrt{\nu^2 + \chi^2}\;
I_0\!\left(\frac{n \beta p_T \chi}{\nu}\right)
\nonumber \\
&\quad \times
K_1\!\left(n \beta m_T \sqrt{1 + \frac{\chi^2}{\nu^2}}\right)
-
\chi\, p_T\;
I_1\!\left(\frac{n \beta p_T \chi}{\nu}\right)
\nonumber \\
&\quad \times
K_0\!\left(n \beta m_T \sqrt{1 + \frac{\chi^2}{\nu^2}}\right)
\Bigg]
\chi\, d\chi,
\label{eq:spectrum}
\end{align}

Here, $g$ denotes the particle degeneracy factor, and $R$ represents the transverse radius of the fireball at freeze-out. The term $\sum_{n=1}^{\infty} \epsilon_n$ accounts for quantum statistical effects, with
\[
\epsilon_n =
\begin{cases}
+1, & \text{for bosons}, \\
(-1)^{n+1}, & \text{for fermions}.
\end{cases}
\]
The transverse mass is defined as $m_T = \sqrt{m^2 + p_T^2}$, where $p_{\rm T}$ is the particle transverse momentum. The variables $\nu$ and $\chi$ are given by $\nu \equiv \tau_{3f}/R$ and $\chi \equiv \rho/R$, where $\rho$ is the radial coordinate in the transverse plane and $\tau_{3f}$ denotes the freeze-out proper time. The functions $I_n(x)$ and $K_n(x)$ correspond to the modified Bessel functions of the first and second kind, respectively. Finally, $\beta \equiv 1/T$ is the inverse temperature, with $T$ being the freeze-out temperature.



The average transverse velocity at freeze-out, derived using the hydrodynamic solutions, is given by

\begin{equation}
\langle v_T \rangle = \sqrt{1 + \nu^2} - \nu^2 \log \left( \frac{\nu}{\sqrt{1 + \nu^2} - 1} \right)
\end{equation}

This expression shows that $\langle v_T \rangle$ decreases monotonically with increasing $\nu$ and approaches unity in the limit $\nu \to 0$. Since $\nu = \tau_{3f}/R$ carries the dimension of inverse velocity, it effectively characterizes the strength of collective expansion at freeze-out.

In the following section, Eq.~(\ref{eq:spectrum}) is employed to describe the core component of the transverse momentum spectra. The key parameters are the temperature $T$ and the expansion parameter $\nu$ (or equivalently $\langle v_T \rangle$). 
The present framework determines the average transverse velocity directly from the underlying hydrodynamic solution.

The high-$p_{\rm T}$ power-law tail observed in experimental particle spectra cannot be described within a purely hydrodynamic framework. Instead, it is commonly modeled using a core–corona picture~\cite{Anton_SHM}. Even in nucleus–nucleus collisions at small impact parameters, a fraction of nucleon–nucleon interactions occurs in the so-called corona region, where the overlap density is significantly lower than the maximum density reached in the collision. In this dilute region, where nucleons undergo on average one or fewer collisions, the formation of a QGP is unlikely; thus, these interactions are treated as proton–proton–like processes. In contrast, the core region corresponds to the high-density zone where thermalization is assumed to occur. The corona is defined as the region where the local density drops to about 10\% of the central nuclear density $\rho_0$, with $\rho_0 \approx 0.16~\mathrm{fm}^{-3}$ for a heavy nucleus at rest. The transverse momentum distribution in $pp$ collisions is parametrized as
\begin{equation}
\frac{d^2\sigma^{pp}}{dy\,dp_T} = C \times \frac{p_T}{\left(1 + \left(\frac{p_T}{p_0}\right)^2 \right)^n}
\label{eq:corona}
\end{equation}

where the parameters $C$, $p_0$, and $n$ are determined by fits to experimental data for each particle species. The normalization is constrained such that the integral of the distribution reproduces the measured $pp$ cross section $d\sigma/dy$. 

\begin{figure}
	\centering 
	\includegraphics[width=0.4\textwidth]{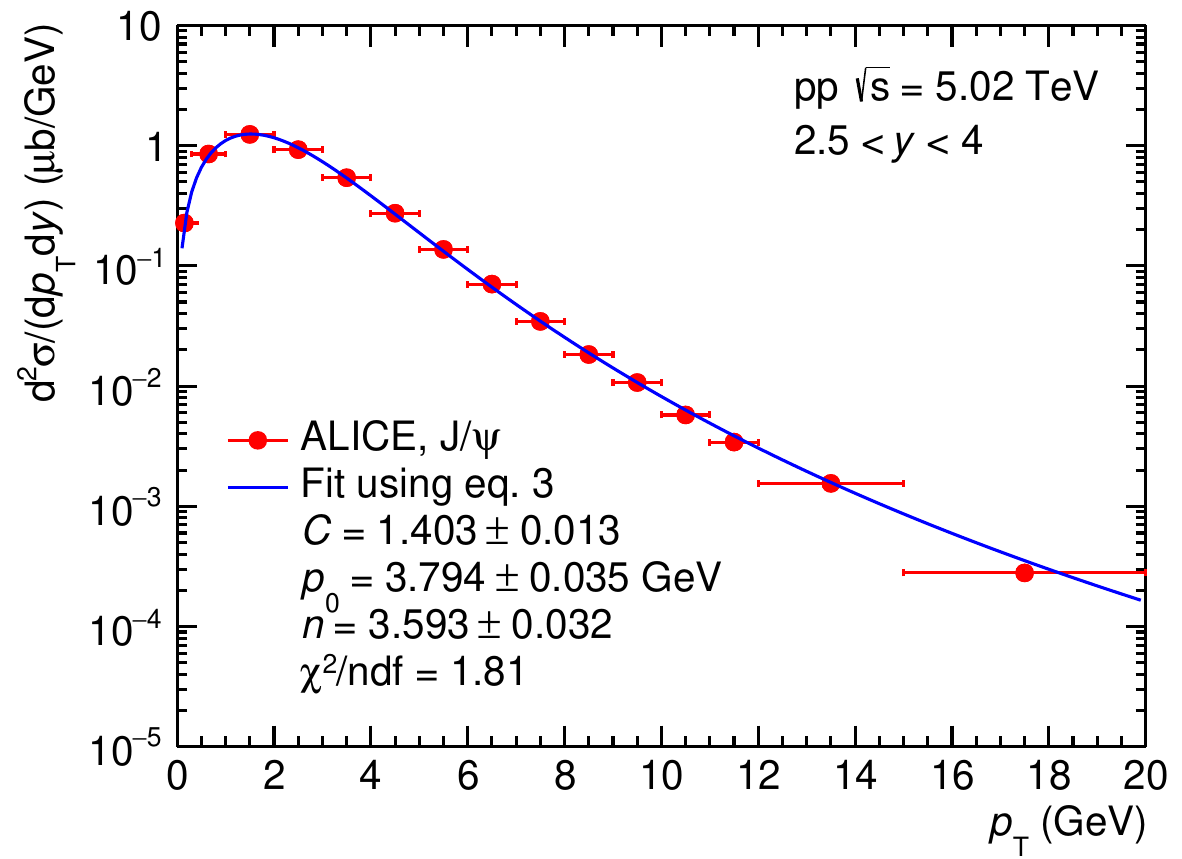}
	\caption{$p_{\rm T}$ spectra of J/$\psi$ in pp collisions at forward rapidity ($2.5 < y < 4$) at $\sqrt{s} = 5.02$ TeV fitted using eq.~\ref{eq:corona}. The data points correspond to measurements from ALICE~\cite{ALICE_pp_JPsi_PsiP_fwdy}, while the solid line represents the fitted curve.} 
	\label{Power_law_JPsi}%
\end{figure}

\begin{figure}
	\centering 
	\includegraphics[width=0.4\textwidth]{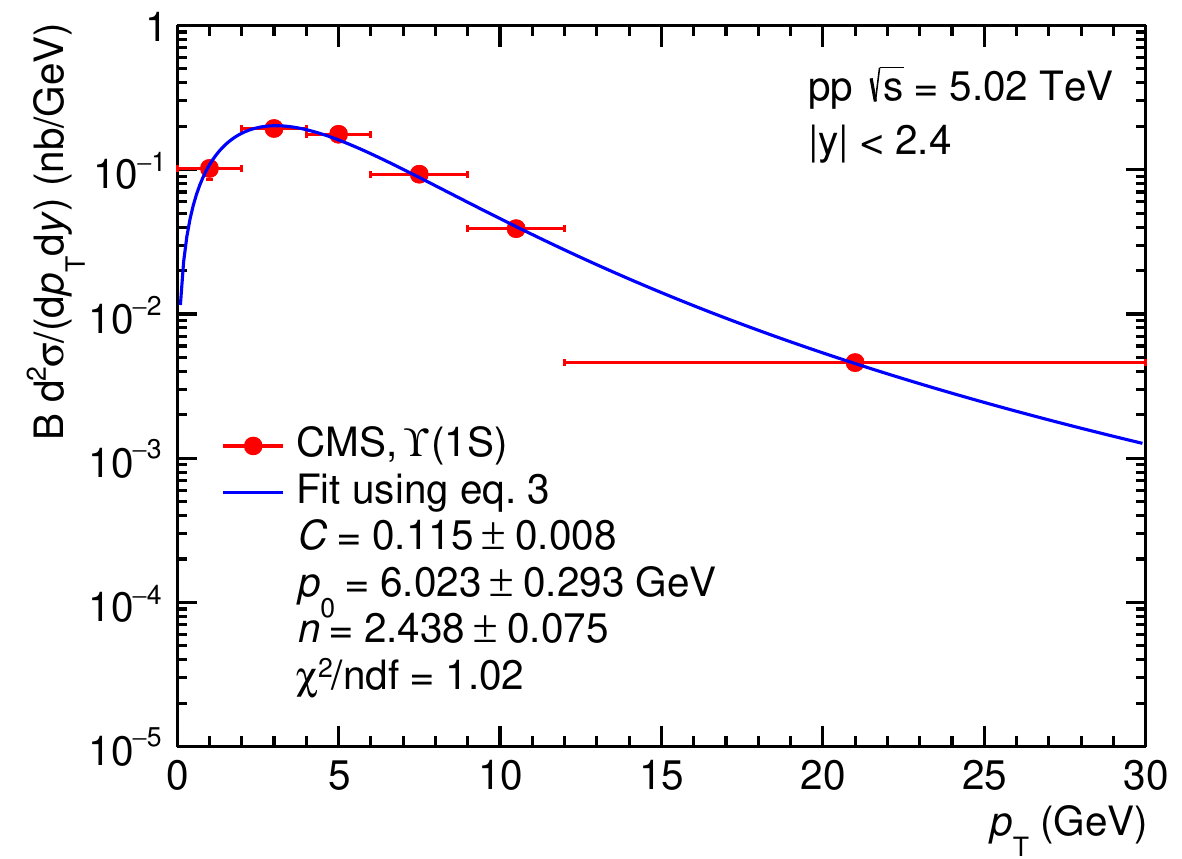}	
	\caption{$p_{\rm T}$ spectra of $\Upsilon(1S)$ in pp collisions at midrapidity ($|\it{y}|<$ 2.4) at $\sqrt{s} = 5.02$ TeV fitted using eq.~\ref{eq:corona}. The data points correspond to measurements from CMS~\cite{CMS_Y1S_midy}, while the solid line represents the fitted curve.} 
	\label{Power_law_UPsi}%
\end{figure}

For J/$\psi$, the $p_{\rm T}$-differential cross sections in $pp$ collisions are taken from ALICE measurements at $\sqrt{s} = 5.02$ TeV, both at midrapidity ($|y| < 0.9$)~\cite{ALICE_pp_JPsi_midy} and forward rapidity ($2.5 < y < 4$)~\cite{ALICE_pp_JPsi_PsiP_fwdy}. The $\psi(2S)$ reference is obtained from ALICE measurements at the same energy and forward rapidity~\cite{ALICE_pp_JPsi_PsiP_fwdy}. For bottomonium states, the $pp$ cross sections of $\Upsilon(1S)$, $\Upsilon(2S)$, and $\Upsilon(3S)$ are taken from CMS measurements at midrapidity ($|y| < 2.4$) at $\sqrt{s} = 5.02$ TeV~\cite{CMS_Y1S_midy}. The results of the fits using Eq.~\ref{eq:corona} for J/$\psi$ and $\Upsilon(1S)$ are shown in Fig.~\ref{Power_law_JPsi} and Fig.~\ref{Power_law_UPsi}, respectively. Similar fits are performed for the other quarkonium states as well. The power-law parameterization reproduces the measured $p_{\rm T}$ spectra remarkably well over the entire kinematic range considered. 
To account for the contribution from the corona in nucleus–nucleus collisions, the $pp$ differential cross section obtained from the fitting is scaled by the nuclear overlap function $T_{AA}^{\mathrm{corona}}$, which represents the number of binary nucleon–nucleon collisions occurring in the corona region.

In summary, the transverse momentum spectra of the considered quarkonia in heavy--ion collisions are constructed by combining the soft component, described by the hydrodynamic framework, with the high-$p_{\rm T}$ component originating from the corona contribution.

\section{Results and discussion}

\begin{figure}
	\centering 
	\includegraphics[width=0.4\textwidth]{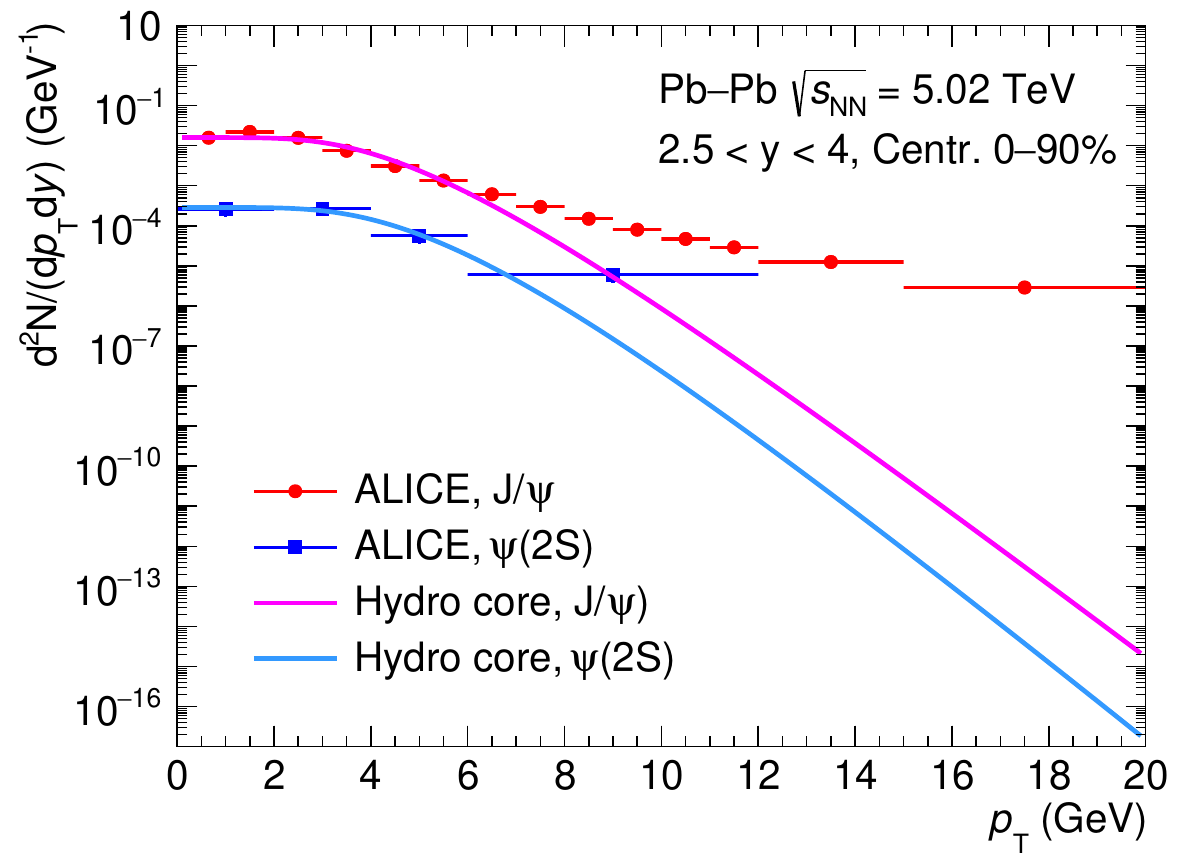}
	\caption{$p_{\rm T}$ spectra of J/$\psi$ and $\psi(2S)$ in Pb--Pb collisions at forward rapidity ($2.5 < y < 4$) for 0--90\% centrality at $\sqrt{s_{\rm NN}} = 5.02$ TeV. The data points correspond to measurements from ALICE~\cite{ALICE_PsiP_PRL}, while the solid lines represent fitted curve from the model calculations (eq.~\ref{eq:spectrum}).} 
	\label{Hydro_core_charmonium}%
\end{figure}

\begin{figure}
	\centering 
	\includegraphics[width=0.4\textwidth]{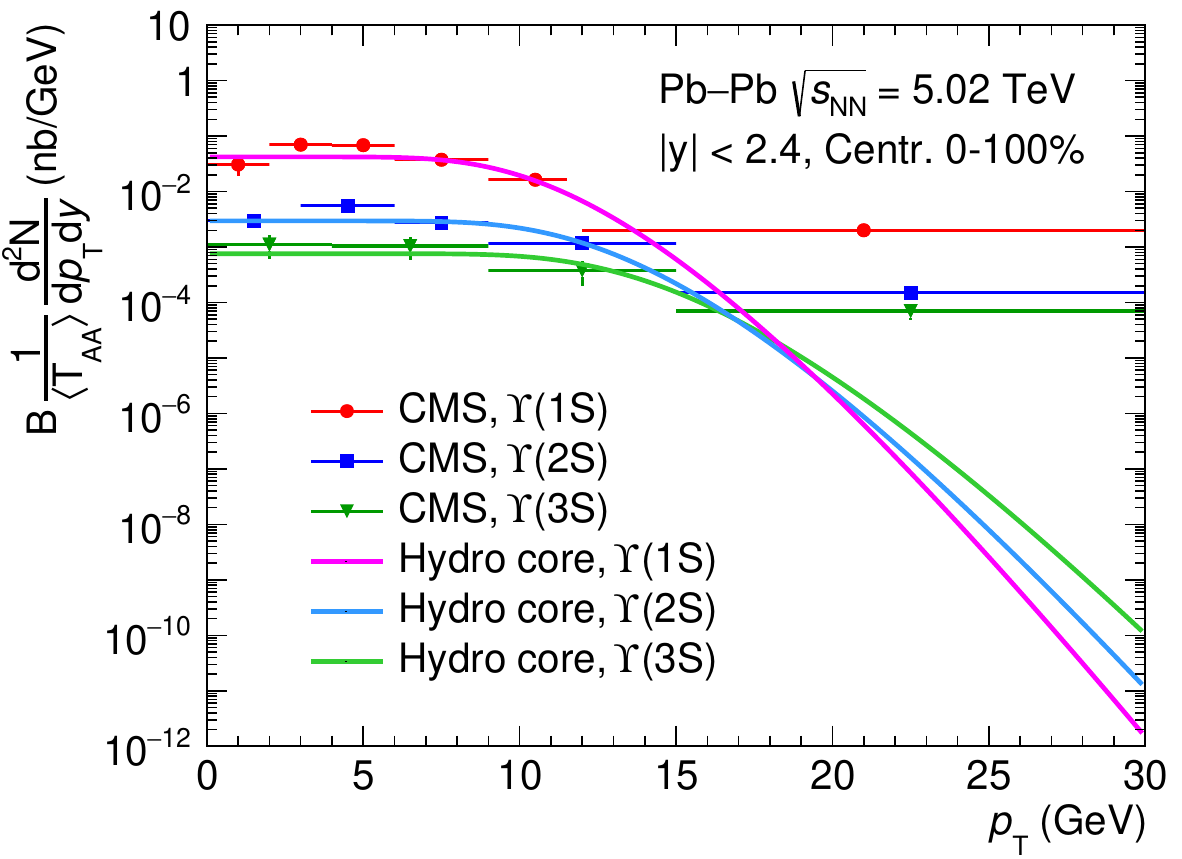}	
	\caption{$p_{\rm T}$ spectra of $\Upsilon(1S)$, $\Upsilon(2S)$, and $\Upsilon(3S)$ in Pb--Pb collisions at midrapidity ($|\it{y}|<$ 2.4) for 0--100\% centrality at $\sqrt{s_{\rm NN}} = 5.02$ TeV. The data points correspond to measurements from CMS~\cite{CMS_Y1S_midy,CMS_Y3S_midy}, while the solid lines represent fitted curve from the model calculations (eq.~\ref{eq:spectrum}).} 
	\label{Hydro_core_bottomonium}%
\end{figure}

The transverse momentum spectra calculated from Eq.~\ref{eq:spectrum} are fitted to the charmonium measurements from ALICE~\cite{ALICE_PsiP_PRL} and bottomonium data from CMS~\cite{CMS_Y1S_midy,CMS_Y3S_midy} in Pb–Pb collisions at $\sqrt{s_{\rm NN}} = 5.02$ TeV. The model parameters are determined via a $\chi^2$ minimization procedure. The simultaneous fit results, shown in Fig.~\ref{Hydro_core_charmonium}, provide a good description of the J/$\psi$ (for $p_{\rm T}$ $<$ 8 GeV) and $\psi(2S)$ spectra, with a freeze-out temperature $T \approx 160$ MeV and an average transverse flow velocity $\langle v_T \rangle \approx 0.60$, with $\chi^2$/ndf of 5.9. Similarly, as shown in Fig.~\ref{Hydro_core_bottomonium}, the $\Upsilon(1S)$, $\Upsilon(2S)$ and $\Upsilon(3S)$ spectra measured by CMS are well described with a freeze-out temperature $T \approx 224$ MeV and an average transverse flow velocity $\langle v_T \rangle \approx 0.56$, with $\chi^2$/ndf = 4.6. These values are consistent with earlier studies~\cite{Gupta,chem_charmonium}. 
A clear distinction between charmonium and bottomonium is reflected in the extracted freeze-out parameters: bottomonium favors a significantly higher freeze-out temperature and a comparatively smaller transverse flow velocity. This suggests that bottomonium decouples earlier from the system and probes the hotter, early-time stages of the medium evolution, whereas charmonium remains sensitive to the later stages, where stronger collective flow develops and medium-driven dynamics play a more prominent role.


It is important to note that the analytical solution provides a transparent baseline for understanding the role of collective expansion and freeze-out geometry in quarkonium production. The framework is expected to be most reliable in the low- and intermediate-$p_{\rm T}$ region, where the assumption of near-local equilibrium remains valid. At higher $p_{\rm T}$, the spectra are increasingly dominated by hard processes and non-equilibrium contributions, including non-prompt production from heavy-flavor decays, medium-modified fragmentation and other perturbative QCD effects, which are not fully captured within the present hydrodynamic description. To account for these contributions, a core--corona framework is employed, where the corona component represents the non-thermal hard-scattering contribution parametrized using the measured $pp$ spectra.

\begin{figure}
	\centering 
	\includegraphics[width=0.4\textwidth]{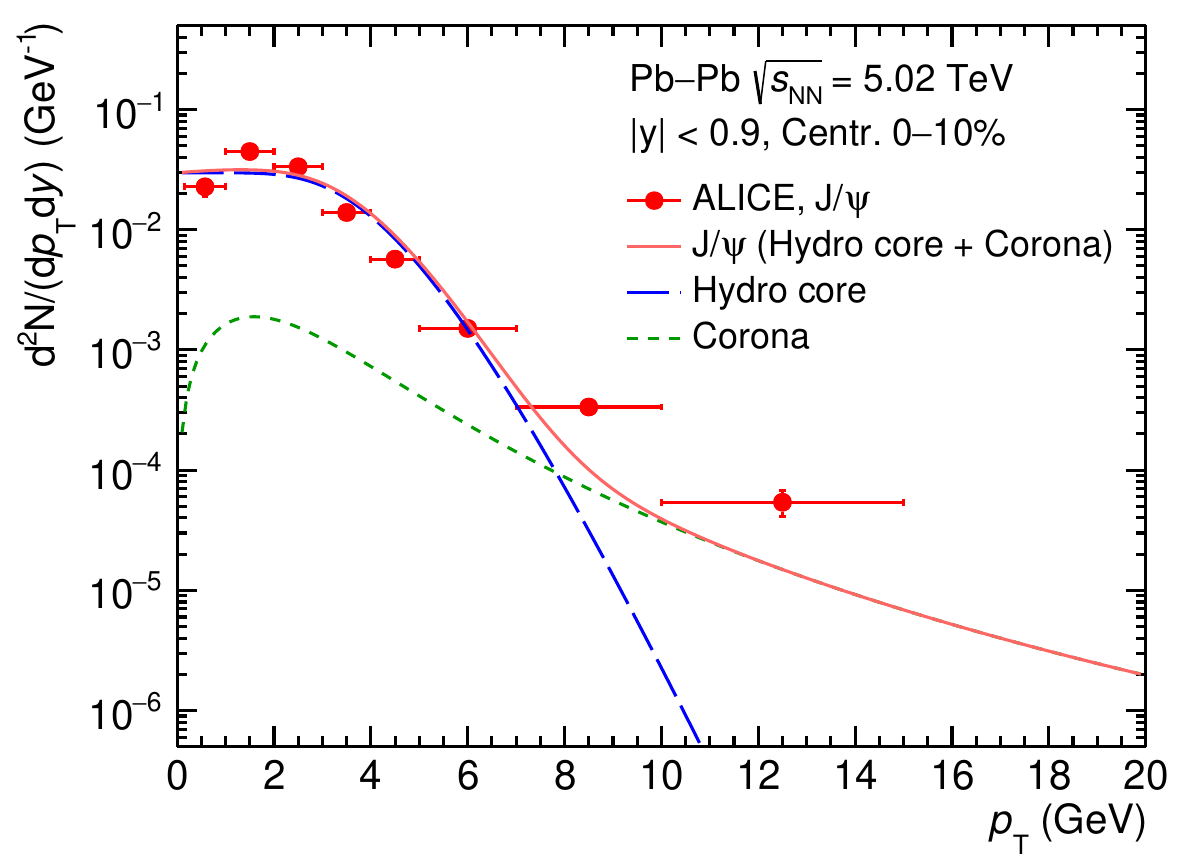}	
	\caption{$p_{\rm T}$ spectra of $J/\psi$ in central (0--10\%) Pb--Pb collisions at midrapidity ($|y| < 0.9$) at $\sqrt{s_{\rm NN}} = 5.02$ TeV, compared with ALICE measurements~\cite{ALICE_JPsi_midy}. In addition to the total spectrum, the individual contributions from the hydrodynamic core component and the corona are also displayed.} 
	\label{Model_ALICE_JPsi_midy}%
\end{figure}

Figure~\ref{Model_ALICE_JPsi_midy} presents the calculated $p_{\rm T}$ spectrum of J/$\psi$ in central (0--10\%) Pb--Pb collisions at midrapidity ($|y| < 0.9$) at $\sqrt{s_{\rm NN}} = 5.02$ TeV, compared with measurements from ALICE~\cite{ALICE_JPsi_midy}. In addition to the total spectrum, the individual contributions from the hydrodynamic core and the corona components are shown to illustrate their relative roles. The model describes the data very well across the full $p_{\rm T}$ range, although it shows a slight underestimation at higher $p_{\rm T}$ ($p_T > 8$ GeV). This discrepancy may suggest the contribution from additional production mechanisms which can enhance yields in the high-$p_{\rm T}$ region.

\begin{figure}
	\centering 
	\includegraphics[width=0.4\textwidth]{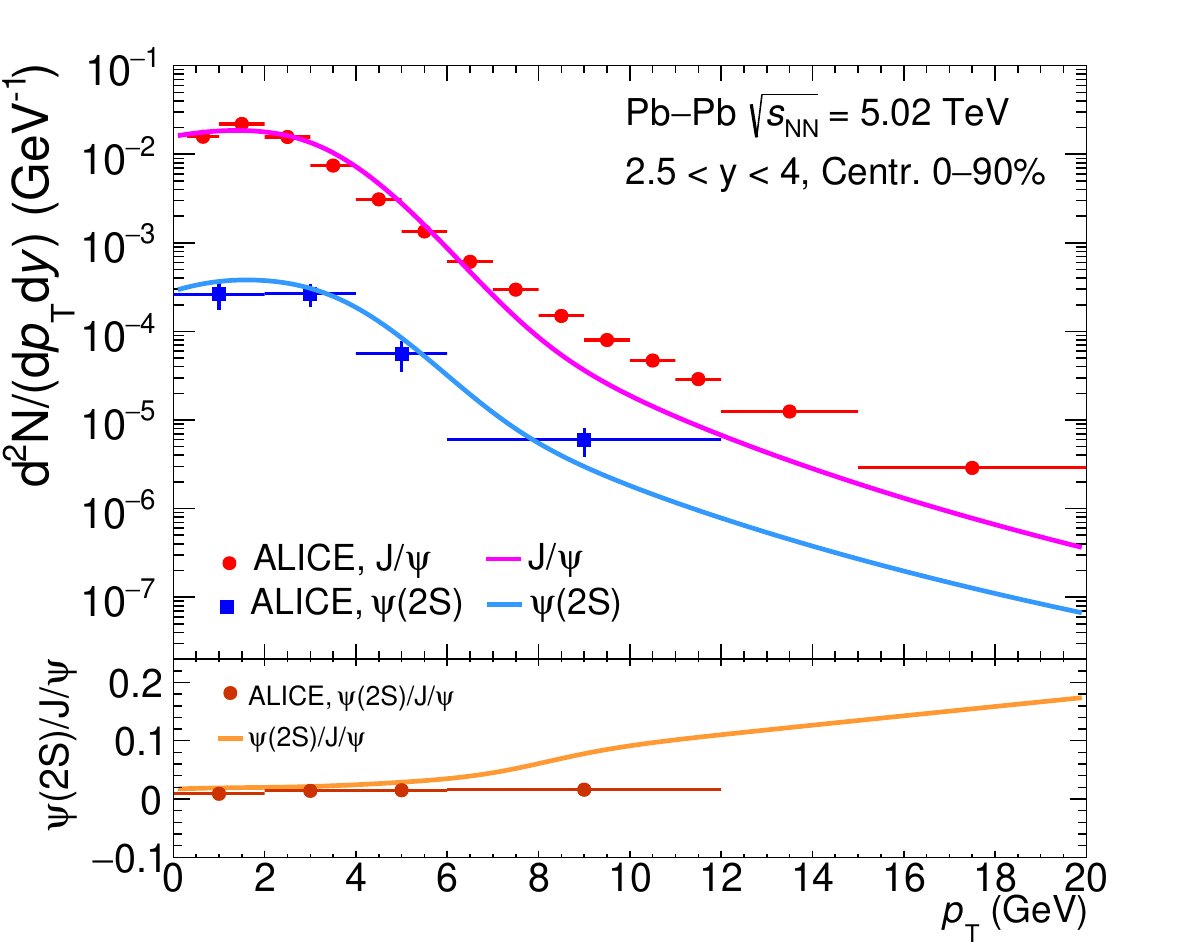}	
	\caption{Upper panel: $p_{\rm T}$ spectra of J/$\psi$ and $\psi(2S)$ in Pb--Pb collisions at forward rapidity ($2.5 < y < 4$) for 0--90\% centrality at $\sqrt{s_{\rm NN}} = 5.02$ TeV, compared with measurements from ALICE~\cite{ALICE_PsiP_PRL} in the same kinematic region. Lower panel: Corresponding $\psi(2S)$/J/$\psi$ yield ratio, compared with the ALICE measurement~\cite{ALICE_PsiP_PRL}. The solid curves represent the total calculated $p_{\rm T}$ spectra, including contributions from both the hydrodynamic core and the corona} 
	\label{Model_ALICE_JPsi_PsiP_fwdy}%
\end{figure}

In the upper panel of Fig.~\ref{Model_ALICE_JPsi_PsiP_fwdy}, the calculated $p_{\rm T}$ spectrum of J/$\psi$ in Pb--Pb collisions at forward rapidity ($2.5 < y < 4$) for 0--90\% centrality at $\sqrt{s_{\rm NN}} = 5.02$ TeV is compared with measurements from the ALICE at forward rapidity~\cite{ALICE_PsiP_PRL}. The model provides a good description of the data at low $p_{\rm T}$, while it underestimates the yield at higher $p_{\rm T}$ ($p_T > 8$ GeV). This deviation may indicate the presence of additional production mechanisms which can enhance yields in the high-$p_{\rm T}$ region.

The predictions for the $\psi(2S)$ state in the same collision system and rapidity interval are also presented and compared with measurements from the ALICE~\cite{ALICE_PsiP_PRL} in the upper panel of Fig.~\ref{Model_ALICE_JPsi_PsiP_fwdy}. The model provides an excellent description of the $\psi(2S)$ data over the entire $p_{\rm T}$ range. The lower panel shows the corresponding $\psi(2S)$/J/$\psi$ production ratio, which is found to be in good agreement with the ALICE measurements within the available $p_{\rm T}$ range. The agreement indicates that collective expansion, along with the corona contribution, captures the essential features of charmonium production in heavy--ion collisions at LHC energies.

\begin{figure}
	\centering 
	\includegraphics[width=0.4\textwidth]{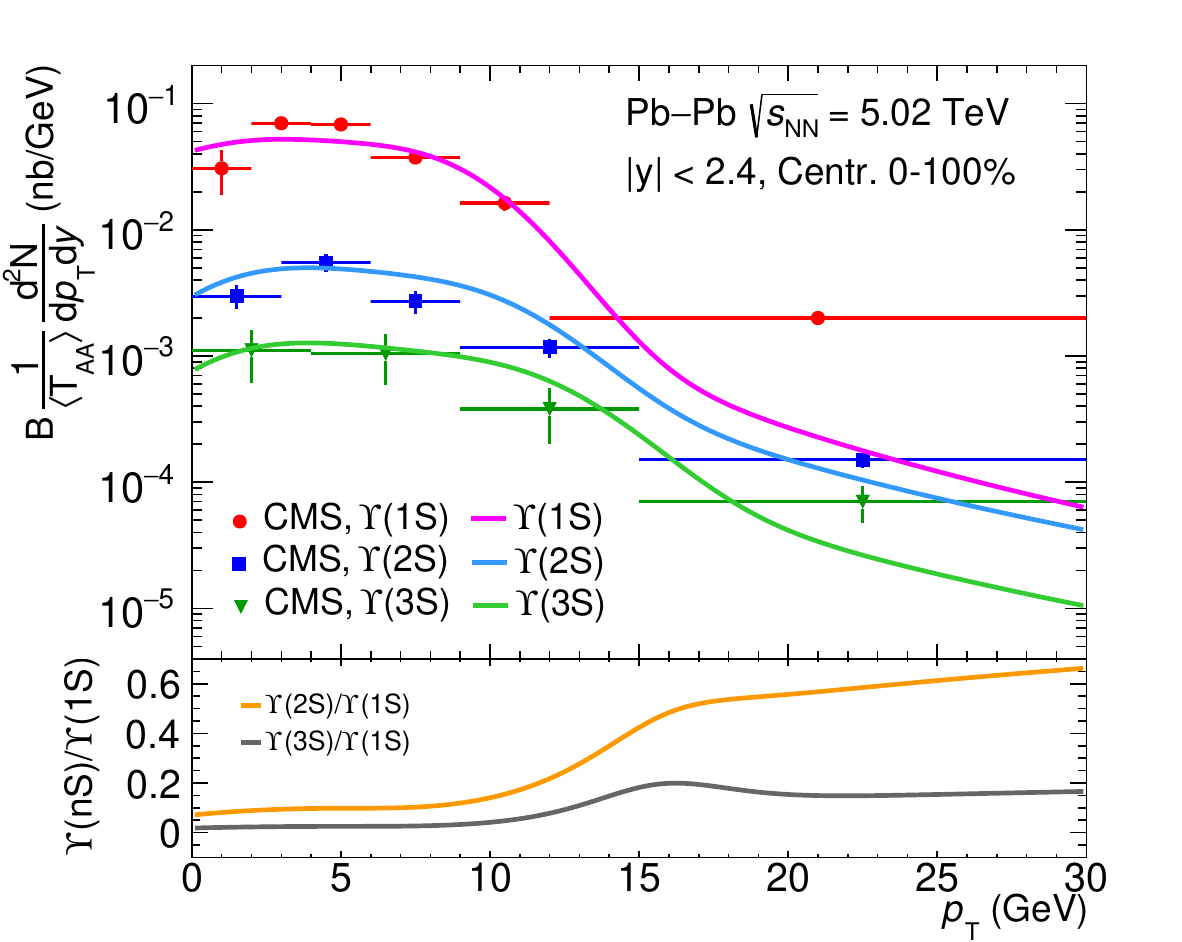}	
	\caption{Upper panel: $p_{\rm T}$ spectra of $\Upsilon(1S)$, $\Upsilon(2S)$, and $\Upsilon(3S)$ in Pb--Pb collisions at midrapidity ($|y| < 2.4$) for 0--100\% centrality at $\sqrt{s_{\rm NN}} = 5.02$ TeV, compared with measurements from CMS~\cite{CMS_Y1S_midy,CMS_Y3S_midy} in the same kinematic region. Lower panel: Corresponding model predictions for the yield ratios $\Upsilon(2S)/\Upsilon(1S)$ and $\Upsilon(3S)/\Upsilon(1S)$. The solid curves represent the total calculated $p_{\rm T}$ spectra, including contributions from both the hydrodynamic core and the corona.} 
	\label{Model_CMS_YnS_midy}%
\end{figure}

Figure~\ref{Model_CMS_YnS_midy} shows the $p_{\rm T}$ spectra of $\Upsilon(1S)$, $\Upsilon(2S)$ and $\Upsilon(3S)$ in Pb--Pb collisions at midrapidity ($|y| < 2.4$) for 0--100\% centrality at $\sqrt{s_{\rm NN}} = 5.02$ TeV, compared with measurements from the CMS~\cite{CMS_Y1S_midy,CMS_Y3S_midy}. The model provides a good description of the data over the measured $p_{\rm T}$ range for all three bottomonium states, indicating that collective flow effects together with the corona contribution are sufficient to describe the bottomonium production in heavy--ion collisions at LHC energies.
The lower panel of Fig.~\ref{Model_CMS_YnS_midy} presents the corresponding yield ratios $\Upsilon(2S)/\Upsilon(1S)$ and $\Upsilon(3S)/\Upsilon(1S)$. The model reproduces the
observed suppression hierarchy among the excited states, consistent with the expected sequential dissociation pattern governed by their binding energies.

\section{Summary}

In this work, we have presented a systematic analysis of the transverse momentum spectra of charmonium (J/$\psi$, $\psi(2S)$) and bottomonium ($\Upsilon(nS)$) states produced in Pb--Pb collisions at $\sqrt{s_{\rm NN}} = 5.02$ TeV within an analytical relativistic hydrodynamics framework. The medium evolution is modeled assuming cylindrical symmetry with boost-invariant longitudinal expansion and Hubble-like transverse flow. The quarkonium spectra are evaluated using the Cooper--Frye prescription on a constant-temperature freeze-out hypersurface, supplemented by a core--corona approach to account for both thermalized and non-thermalized production components.

A notable difference between charmonium and bottomonium emerges in the extracted freeze-out parameters, with bottomonium favoring a higher freeze-out temperature and slightly lower transverse flow velocity. This indicates that bottomonium states predominantly probe earlier and hotter stages of the medium evolution, while charmonium production is more influenced by later-stage dynamics and possible regeneration processes.

For charmonium, the model provides a good description of the J/$\psi$ and $\psi(2S)$ spectra measured by the ALICE collaboration over a wide $p_{\rm T}$ range. The $\psi(2S)$/J/$\psi$ ratio is also well reproduced, indicating that the combined effects of collective flow and corona contributions capture the essential features of charmonium production. However, the observed deviation at high $p_{\rm T}$ in the J/$\psi$ spectrum suggests the presence of additional hard production mechanisms which are not explicitly included in the present framework.

In the bottomonium sector, the calculated $p_{\rm T}$ spectra of $\Upsilon(1S)$, $\Upsilon(2S)$, and $\Upsilon(3S)$ show good agreement with measurements from the CMS collaboration. The corresponding yield ratios are also well described, reproducing the expected sequential suppression pattern, where more weakly bound states exhibit stronger suppression. This behavior reflects the sensitivity of bottomonium states to the medium temperature and their binding energies, with minimal contributions from regeneration effects compared to the charmonium sector.


Overall, the model provides a good description of the data in the low- and intermediate-$p_{\rm T}$ regions, where collective expansion dominates. At higher $p_{\rm T}$, the hydrodynamic contribution alone underestimates the data, reflecting the limitations of the ideal hydrodynamic description and the growing importance of non-thermal processes. The inclusion of the corona component improves the agreement in this region, although residual deviations, specially for J/$\psi$, may indicate the need for a more complete treatment including microscopic production mechanisms.

These results highlight the complementary roles of charmonium and bottomonium as probes of the quark--gluon plasma and emphasize the importance of combining analytical and phenomenological approaches to gain deeper insight into the properties of the strongly interacting medium created at the LHC.

\section*{Acknowledgements}

It is a pleasure to thank Amaresh Jaiswal and Anton Andronic for helpful discussions.


\begin{thebibliography}{50}

\bibitem{Matsui} T. Matsui and H. Satz, Phys. Lett. B {\bf 178}, 416 (1986).

\bibitem{Rvogt} R. Vogt, Phys. Rep. {\bf 310}, 197 (1999).

\bibitem{Satz} H. Satz, J. Phys. G {\bf 32}, R25 (2006).

\bibitem{Kluberg} L. Kluberg and H. Satz, Relativistic Heavy Ion Physics Vol. {\bf 23} (Springer-Verlag, Berlin, Heidelberg,
2010).

\bibitem{Rapp} R. Rapp, D. Blaschke, and P. Crochet, Prog. Part. Nucl. Phys. {\bf 65}, 209 (2010).

\bibitem{Zhao} J. Zhao, K. Zhou, S. Chen, and P. Zhuang, Prog. Part. Nucl. Phys. {\bf 114}, 103801 (2020).

\bibitem{Mocsy} A. Mocsy, P. Petreczky, and M. Strickland, Int. J. Mod. Phys. A {\bf 28}, 1340012 (2013).

\bibitem{Rothkopf} A. Rothkopf, Heavy quarkonium in extreme conditions, Phys.
Rep. {\bf 858}, 1 (2020).

\bibitem{Karsch} F. Karsch, D. Kharzeev, and H. Satz, Phys. Lett. B {\bf 637}, 75 (2006).

\bibitem{Bhaduri} P. P. Bhaduri, P. Hegde, H. Satz, and P. Tribedy, Lecture Notes in Physics Vol. {\bf 785} (Springer,
Berlin, 2010), p. 179.
\bibitem{Abreu} M. C. Abreu {\it et al.} (NA50 Collaboration), Phys. Lett. B {\bf 477}, 28 (2000).

\bibitem{Arnaldi} R. Arnaldi {\it et al.} (NA60 Collaboration), Phys. Rev. Lett {\bf 99}, 132032 (2007).

\bibitem{Adare1} A. Adare {\it et al.} (PHENIX Collaboration), Phys. Rev. Lett. {\bf 98}, 232301 (2007).

\bibitem{Adare2} A. Adare {\it et al.} (PHENIX Collaboration), Phys. Rev. C {\bf 84}, 054912 (2011).

\bibitem{Adam} J. Adam {\it et al.} (STAR Collaboration), Phys. Lett. B {\bf 797}, 134917 (2019).

\bibitem{Abelev} B. B. Abelev {\it et al.} (ALICE Collaboration), Phys. Lett. B {\bf 734}, 314 (2014).

\bibitem{Adam2} J. Adam {\it et al.} (ALICE Collaboration), J. High Energy Phys. {\bf 05} (2016) 179.

\bibitem{Adam3} J. Adam {\it et al.} (ALICE Collaboration), Phys. Lett. B {\bf 766}, 212 (2017).

\bibitem{Alice_quarkonia} S. Acharya {\it et al.} (ALICE Collaboration), J. High Energy Phys.
{\bf 02}, 41 (2020).

\bibitem{ALICE_JPsi_midy} S. Acharya {\it et al.} (ALICE Collaboration), Phys. Lett. B {\bf 849}, 138451 (2024).

\bibitem{ALICE_PsiP_PRL} S. Acharya {\it et al.} (ALICE Collaboration), Phys. Rev. Lett. {\bf 132}, 042301 (2024).

\bibitem{CMS} S. Chatrchyan {\it et al.} (CMS Collaboration), J. High Energy Phys. {\bf 05}, 063 (2012).

\bibitem{CMS_PsiP} A. M. Sirunyan {\it et al.} (CMS Collaboration), Phys. Rev. Lett. {\bf 118}, 162301 (2017).

\bibitem{CMS_Prompt} A. M. Sirunyan {\it et al.} (CMS Collaboration), Eur.
Phys. J. C {\bf 78}, 509 (2018).

\bibitem{Atlas_quarkonia} M. Aaboud {\it et al.} (ATLAS Collaboration), Eur. Phys. J. C {\bf 78}, 762
(2018).

 \bibitem{LHCb_quarkonia} R. Aaij {\it et al.} (LHCb Collaboration), Phys. Rev. C {\bf 105}, L032201 (2022).
\bibitem{ALICE_Y1S_fwdy} S. Acharya {\it et al.} (ALICE Collaboration), Phys. Lett. B {\bf 822}, 136579 (2021).

\bibitem{CMS_Y1S_midy} A. Tumasyan {\it et al.} (CMS Collaboration), Phys. Lett. B {\bf 790}, 270--293 (2019). 

\bibitem{CMS_Y2S_midy} A. Tumasyan {\it et al.} (CMS Collaboration), Phys. Rev. Lett. {\bf 120}, 142301 (2018).

\bibitem{CMS_Y3S_midy} A. Tumasyan {\it et al.} (CMS Collaboration), Phys. Rev. Lett. {\bf 133}, 022302 (2024). 

\bibitem{ATLAS_YnS_midy} G. Aad {\it et al.} (ATLAS Collaboration), Phys. Rev. C {\bf 107}, 054912 (2023)

\bibitem{Du} X. Du, M. He, and R. Rapp, Phys. Rev. C {\bf 96}, 054901 (2017).

\bibitem{Gazdzicki} M. Gazdzicki and M. I. Gorenstein, Phys. Rev. Lett. {\bf 83}, 4009 (1999).

\bibitem{Bugaev} K. A. Bugaev, M. Gazdzicki, and M. I. Gorenstein, Phys. Lett. B {\bf 544}, 127 (2002).

\bibitem{Anton_SHM} A. Andronic, P. Braun-Munzinger, M. K. Köhler, K. Redlich,
and J. Stachel, Phys. Lett.
B {\bf 797}, 134836 (2019).

\bibitem{CMS_YnS} V. Khachatryan {\it et al.} (CMS Collaboration), Phys. Lett. B {\bf 770}, 357 (2017).

\bibitem{Reygers} K. Reygers, A. Schmah, A. Berdnikova, and X. Sun, Phys. Rev. C {\bf 101}, 064905 (2020).

\bibitem{ALICE_v2_JPsi} S. Acharya {\it et al.}, (ALICE Collaboration), Phys.
Rev. Lett. {\bf 123}, 192301 (2019).

\bibitem{CMS_v2_JPsi} A. M. Sirunyan {\it et al.} (CMS Collaboration), Phys. Lett. B {\bf 819}, 136385 (2021).

\bibitem{ALICE_v2_JPsi2} S. Acharya {\it et al.}, (ALICE Collaboration), Phys. Rev. Lett. {\bf 119}, 242301 (2017).

\bibitem{ALICE_v2_JPsi3} S. Acharya {\it et al.}, (ALICE Collaboration), J. High Energy Phys. {\bf 02}, 12 (2019).

\bibitem{ALICE_v2_JPsi4} S. Acharya {\it et al.}, (ALICE Collaboration), J. High Energy Phys. {\bf 10}, 141 (2020).

\bibitem{Rapp_JPsi} M. He, B. Wu, and R. Rapp, Phys. Rev. Lett. {\bf 128}, 162301 (2022).

\bibitem{Bhaduri_Y} P. P. Bhaduri, N. Borghini, A. Jaiswal, and M. Strickland,
Phys. Rev. C {\bf 100}, 051901(R) (2019).

\bibitem{Bhaduri_Y2} P. P. Bhaduri, M. Alqahtani, N. Borghini, A. Jaiswal, and M. Strickland, Eur. Phys. J. C {\bf 81}, 585 (2021).

\bibitem{Anton_SHM2} A. Andronic, P. Braun-Munzinger, K. Redlich, and J. Stachel, Nucl. Phys. A {\bf 789}, 334 (2007).

\bibitem{Anton_SHM3} A. Andronic, P. Braun-Munzinger, K. Redlich, and J. Stachel, Phys. Lett. B {\bf 652}, 259 (2007).

\bibitem{Anton_SHM4} A. Andronic, P. Braun-Munzinger, M. K. Köhler, and J. Stachel, Nucl. Phys. A {\bf 982}, 759 (2019).

\bibitem{Anton_SHM5} A. Andronic, P. Braun-Munzinger, K. Redlich, and J. Stachel, Nature (London) {\bf 561}, 321 (2018).

\bibitem{Gupta} S. Gupta and R. Sharma, Phys. Rev. C {\bf 89}, 057901 (2014).

\bibitem{Gupta2} S. Chatterjee, R. M. Godbole, and S. Gupta, Phys. Lett. B {\bf 727}, 554 (2013).

\bibitem{Bhaduri_T} D. Kumar, N. Sarkar, P. P. Bhaduri and A. Jaiswal, Phys. Rev. C {\bf 107}, 064906 (2023)

\bibitem{Hydro2} T. Csorgo, F. Grassi, Y.Hama, T.Kodama, Phys. Lett. B {\bf 565}, 107–115 (2003).

\bibitem{Hydro3} S. Lin, J. Liao, Nucl. Phys. A 8{\bf 37}, 195-209 (2010).

\bibitem{Hydro_core} M. S. Ali, D. Biswas, A. Jaiswal, S. K. Singh, Eur. Phys. J. C {\bf 85}, 30 (2025).

\bibitem{GM} M.L. Miller, K. Reygers, S.J. Sanders, P. Steinberg, Annu. Rev. Nucl. Part. Sci. {\bf 57}, 205 (2007).

\bibitem{ALICE_pp_JPsi_midy} S. Acharya {\it et al.} (ALICE Collaboration), J. High Energy Phys. {\bf 10}, 084
(2019).

\bibitem{ALICE_pp_JPsi_PsiP_fwdy} S. Acharya {\it et al.} (ALICE Collaboration), Eur. Phys. J. C {\bf 83}, 61 (2023).

\bibitem{chem_charmonium} A. Andronic {\it et al.}, J. High Energy Phys. {\bf 10}, 229 (2024).




\end{thebibliography}
\end{document}